# Accepted Manuscript

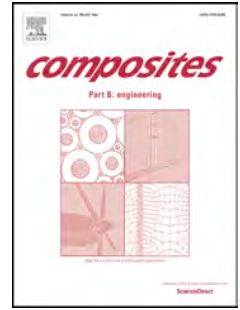

Electrical conductivity of carbon nanofiber reinforced resins: Potentiality of Tunneling Atomic Force Microscopy (TUNA) technique

Marialuigia Raimondo, Liberata Guadagno, Luigi Vertuccio, Carlo Naddeo, Giuseppina Barra, Giovanni Spinelli, Patrizia Lamberti, Vincenzo Tucci, Khalid Lafdi

Please cite this article as: Raimondo M, Guadagno L, Vertuccio L, Naddeo C, Barra G, Spinelli G, Lamberti P, Tucci V, Lafdi K, Electrical conductivity of carbon nanofiber reinforced resins: Potentiality of Tunneling Atomic Force Microscopy (TUNA) technique, *Composites Part B* (2018), doi: 10.1016/ j.compositesb.2018.02.005.

# Electrical conductivity of carbon nanofiber reinforced resins: potentiality of Tunneling Atomic Force Microscopy (TUNA) technique

Marialuigia Raimondo[1,]*, Liberata Guadagno[1], Luigi Vertuccio[1],
Carlo Naddeo[1], Giuseppina Barra[1], Giovanni Spinelli[2],
Patrizia Lamberti[2], Vincenzo Tucci[2], Khalid Lafdi[3]

*[1]Department of Industrial Engineering, University of Salerno,
Via Giovanni Paolo II, 132, 84084, Fisciano (SA), Italy*

*[2]Department of Information and Electrical Engineering and Applied Mathematics,
University of Salerno, Via Giovanni Paolo II, 132, 84084, Fisciano (SA), Italy*

*[3]University of Dayton, 300 College Park, Dayton Ohio, 45440, USA*

*\*Corresponding author*

*Marialuigia Raimondo - e-mail: mraimondo@unisa.it*

**ABSTRACT**

Epoxy nanocomposites able to meet pressing industrial requirements in the field of structural material have been developed and characterized. Tunneling Atomic Force Microscopy (TUNA), which is able to detect ultra-low currents ranging from 80 fA to 120 pA, was used to correlate the local topography with electrical properties of tetraglycidyl methylene dianiline (TGMDA) epoxy nanocomposites at low concentration of carbon nanofibers (CNFs) ranging from 0.05% up to 2% by wt. The results show the unique capability of TUNA technique in identifying conductive pathways in CNF/resins even without modifying the morphology with usual treatments employed to create electrical contacts to the ground.

## 1. Introduction

Epoxy matrix-based composite materials filled with MWCNTs or various carbon additives, such as natural, artificial and exfoliated graphites (EGs), activated carbons, and thick graphene exhibit many properties of great interest both from a scientific and industrial point of view such as low electrical percolation threshold (below 1.5 wt.%) as well as high electromagnetic interference (EMI) shielding capacity, good adhesive properties etc… [1].

CNF-reinforced polymer composites have opened up new perspectives for multifunctional materials. In particular, carbon nanofibers play a promising role due to their potential applications in order to improve mechanical, electrical, thermal performance and relative ease of processing in composites used in the aerospace field [2-7]. Moreover, in high performance resin formulations, CNFs can be used as effective low-cost replacements for carbon nanotubes (CNTs) since the cost of CNFs is much lower than that of CNTs while possessing similar physical properties. The tetrafunctional epoxy resin tetraglycidyl methylene dianiline (TGMDA) cured with the aromatic diamine 4,4'-diaminodiphenylsulfone (DDS) is one of the most widely employed matrices for the production of high performance fiber composites in the aircraft and spacecraft industries [8]. Carbon nanofibers with extremely high aspect ratios combined with low density possess high electrical conductivity. They are excellent nanofiller materials for transforming electrically non-conducting polymers into conductive materials which have a wide range of applications, namely in electromagnetic interference (EMI) shielding (which has become a critical issue in the last several years with the development and advancement of electrical devices), photovoltaic devices, transparent conductive coatings as well as electro-actuating the shape memory polymer composites [9-18] and others [19-21]. Therefore, they are highly sought-after candidate materials to replace traditional metallic materials for lightning strike protection of aircrafts. Modern aircrafts are made of advanced composite materials, which are significantly less conductive than aluminum. Lightning is initiated at the airplane's leading edges, which ionize, creating a strike opportunity. Lightning currents travel along the airplane and exit to the ground, forming a circuit with the airplane between the cloud energy and the ground [22]. Being struck by lightning, the composites are damaged quickly due to the temperature rise from the resistive heating and serious degradation at the lightning attachment points. Therefore, the principle of lightning strike protection is to provide a safe conductive path on the exterior skin. With highly conductive skins, most of the lightning current remains on the exterior skin, without serious damage to the aircraft. Hence, it is necessary to develop advanced composites with high electrical conductivity [6,23-29]. In this regard, the development of polymer composite with nanoscaled modifiers has become an attractive new subject in materials science from researchers worldwide. Epoxy nanocomposites with different CNF loadings to impart electron conduction in epoxy resin, prepared in the lab by means of ultrasonic dispersion, have been investigated in this work. Epoxy nanocomposites with carbon nanofibers, as conductive reinforcement, are characterized by the presence of conductive paths at a relatively very low nanofiber content, determining an increase of conductivity beyond the electrical percolation threshold (EPT) due to an effective connection among nanofibers. It is known that the final conductivity of the polymer composites depends not only on the concentration and on specific conductivity of the nanofiller, but also on their state of dispersion in the matrix. The present study focused on the effect of CNF weight fraction on the conductive behavior of epoxy/CNF composites. It has been demonstrated that the EPT is directly related to the dispersion of the fillers in the polymer matrix, and to overcome the criticality of their agglomeration still remains a challenge [30]. The distribution of particles (e.g. micro and nanoinclusions, carbon nanotubes, carbon nanofibers) within a composite material plays a vital role in dictating many important constitutive properties of the composite in order to determine optimum degree of dispersion for achieving maximum performance [31]. In this work, the drawback of the nanofiller dispersion that is a very critical step prior to the epoxy cure has been successfully overcome by dispersing CNFs into the epoxy liquid mixture thanks to the use of the reactive diluent 1,4-butandiol diglycidyl ether (BDE) into the TGMDA liquid epoxy precursor. In this regard, the reactive diluent BDE inside the unfilled and nanofilled epoxy precursor based on TGMDA has proven to be a key in reducing the viscosity of the epoxy matrix [2-5,8,32-36]. Surface analysis methods have contributed to recent advancements in science, technology and industrial development [37,38]. The development of the scanning tunneling microscopy (STM) revolutionized the study of nanoscale and atomic scale surface structures and properties [39], most of all in the case of electrically conductive nanofillers embedded in matrix [36]. In fact, the need for electrical characterization of surfaces on the nanometer scale in order to improve local conductivity measurements has led quickly to a variety of scanning probe microscopy based techniques. For this type of measurements, usually, two different setups performed in contact mode are used: conductive atomic force microscope (C-AFM) [40-43] and tunneling AFM (TUNA) [36,44,45] depending on the range of currents involved. The first is used to measure current in the range of sub-nA to μA (in particular, higher currents can be measured ranging from 1 pA to 1 μA), the latter for the range between sub-pA to nA (in particular, ultra-low currents (<1pA) ranging from 80 fA to 120 pA can be measured). In this work, TUNA, which utilizes a conductive probe during the measurement process, was used. For the first time, this paper presents a conductivity mapping, performed by TUNA, of CNF reinforced epoxy resin. This novel technology allows to simultaneously map the topography and conductivity of advanced material by applying controlled, low forces on the tip during imaging, which allows a direct comparison between the morphology and the electrical properties at the nanoscale [44,46-49]. For this technique, that uses a conductive AFM probe in contact mode, the sensor signal is the electric current between the AFM tip and the conductive sample for an applied DC bias. This non-contact technique helps in carrying out various non-destructive measurements on electrical conductive nanoparticles to obtain point measurement scan of the sample topography and its corresponding electrical data. It is worth pointing out that there are currently few works dealing with morphological characterizations performed by this new technique. This paper focus on electrical characterization at nanoscale level using Tunneling AFM (TUNA) of carbon nanofiber reinforced epoxy resins. TUNA has proven to be a very effective means of investigation to identify conductive pathways and interconnections in CNF/resins, without undergoing the sample to any prior treatment with silver paint, which is usually employed to create electrical contacts to the ground. In this paper, the morphological characterization is shown together spectroscopic, thermal and mechanical characterizations to preliminary obtain a concise overview of the applicability potential fields. Thermal characterization, performed by DSC analysis, was used to evaluate the curing degree of the developed nanocomposites; whereas Dynamic Mechanical Analysis (DMA) was used to analyze the effect of nanofiller on the storage moduli and the loss factors (tanδ) of the developed nanocomposites. The high performance in mechanical properties strictly related to the high curing degree (almost 100%) reached with the chosen formulation and curing treatment, together with the values in the electrical conductivity and the state of the nanofiller dispersion and interconnections (analyzed by TUNA technique), highlight an interesting applicative potential for the formulated material. FT/IR results were found in good agreement with DSC data. The progressive decrease of the epoxy stretching frequency (906 cm$^{-1}$) was found strictly related to the curing degree evaluated by DSC investigation.

## 2. Experimental section

### 2.1. Materials and epoxy samples preparation

The epoxy matrix T20B was prepared by mixing an epoxy precursor, tetraglycidyl methylene dianiline (TGMDA) (epoxy equivalent weight 117–133 g/eq), with an epoxy reactive diluent 1-4 Butanediol diglycidyl ether (BDE) at a concentration of 80%:20% (by wt), respectively. The unfilled epoxy resin T20BD was obtained after adding to the T20B epoxy mixture, the 4,4'-diaminodiphenyl sulfone (DDS) curing agent at a stoichiometric concentration with respect to all the epoxy rings (TGMDA and BDE). Fig. 1 shows the chemical structures of the compounds used for the epoxy matrix T20BD.

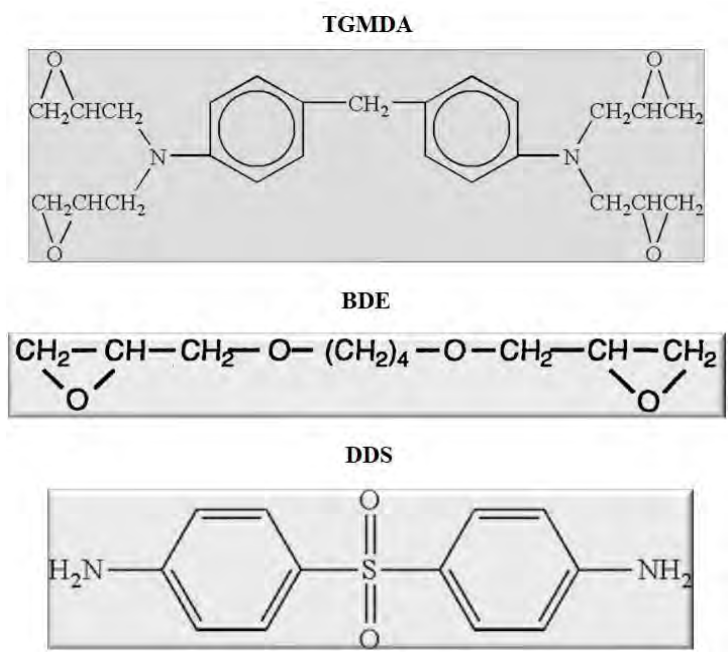

**Fig.1.** Chemical structures of compounds used for the epoxy matrix T20BD.

As already mentioned above, the reactive diluent BDE has proven to be of benefit for thermosetting resins filled with nanoparticles which tend to dramatically increase the viscosity of the resin. It increases the mobility of reactive groups resulting in a higher cure degree than the epoxy precursor alone. This effect is particularly advantageous for nanofilled resins where higher temperature treatments are needed, compared to the unfilled resin, to reach the same cure degree. In this paper, the conductive carbon nanofibers (CNFs) were embedded in the epoxy matrix with the aim of improving the physical properties of the resin. CNFs in the form of powders used in this study were produced at Applied Sciences Inc. and were from the Pyrograf III family. The CNFs, named in this work CNF2500, are obtained by heat treatment at T=2,500°C starting from the pristine carbon nanofibers PR25XTPS1100 where XT indicates the debulked form of the PR25 family, PS indicates the grade produced by pyrolytically stripping the as-produced fiber to remove polyaromatic hydrocarbons from the fiber surface and 1100 is the temperature in the process production. The nanofibers are characterized by (a) an average bulk density of product (g cm$^{-3}$) ranging from 0.0192 to 0.0480; (b) a nanofiber density (including hollow core) (g cm$^{-3}$) from 1.4 to 1.6; (c) a nanofiber wall density (g cm$^{-3}$) from 2.0 to 2.1; (d) an average catalyst (iron) content (ppm) <14 000; (e) an average outer diameter (nm) from 125 to 150; (f) an average inner diameter (nm) from 50 to 70; (g) an average specific surface area, m$^2$ g$^{-1}$ from 65 to 75; (h) a total pore volume (cm$^3$ g$^{-1}$) of 0.140; (i) an average pore diameter (angstroms Å) of 82.06 and lengths ranging from 50 to 100 µm. The PR25XTPS1100 carbon nanofibers were heat-treated to 2500°C to provide the best combination of mechanical and electrical properties [4]. The heat treatment was performed in an atmosphere controlled batch furnace. Approximately 300 g of nanofibers were placed in a ceramic crucible for the heat treatment. The furnace was purged with nitrogen gas for 1 h prior to heating. The heating rate was 100°C h$^{-1}$ and the furnace was held at a temperature of 2500°C for 1 h prior to cooling. Epoxy blend and DDS were mixed at 120°C and the CNF2500 were added and incorporated into the matrix by using an ultrasonication for 20 min. An ultrasonic device, Hielscher model UP200S (200 W, 24 kHz) was used. CNF2500 were dispersed within the epoxy resin at loading rates of 0.05, 0.32, 0.64, 0.8, 1.00, 1.3 and 2% by weight. Epoxy nanocomposites with loads of CNF2500 beyond 1.3% by weight have difficulty in establishing a homogeneous mixture [35,50]. All the epoxy mixtures were cured by a two-stage curing cycle: a first isothermal stage was carried out at the lower temperature of 125°C for 1 h and the second isothermal stage at higher temperatures up to 200°C for 3 h. This curing cycle was chosen because it meets industrial requirements to manufacture the carbon fiber reinforced composites CFRCs (the temperature/time of the first step is lower than the second one to facilitate the CF impregnation before the resin solidification). In this work, the samples have been named with the following acronyms: T20BD for the unfilled epoxy formulation, T20BD+(%)CNF2500 for the nanofilled epoxy resins where (%) represents the different CNF2500 load percentage with which all the nanocomposites analysed in this work have been obtained.

*2.2. Methods*

*2.2.1. Dynamic mechanical analysis (DMA)*

Dynamic mechanical properties of the unfilled epoxy matrix T20BD and CNF-reinforced epoxy nanocomposites were performed with a dynamic mechanical thermo-analyzer (Tritec 2000 DMA -Triton Technology). Solid samples with dimensions 2 x 10 x 35 mm$^3$ were tested by applying a variable flexural deformation in three points bending mode. The displacement amplitude was set to 0.03 mm, whereas the measurements were performed at the frequency of 1 Hz. The range of temperature was from -90°C to 310°C at the scanning rate of 3°C/min.

### 2.2.2. Fourier Transform Infrared Spectroscopy (FTIR) Analysis

The infrared spectra were obtained in absorbance by using a Bruker Vertex 70 FTIR-spectrophotometer with a resolution of 2 cm$^{-1}$ (32 scans collected) in the range of 4000-400 cm$^{-1}$, and the samples were spread on the KBr slice.

### 2.2.3. Differential Scanning Calorimetry (DSC) Analysis

Thermal analysis was performed with a Mettler DSC 822 differential scanning calorimeter in a flowing nitrogen atmosphere. The samples were analyzed in the temperature range of 0°C-300°C with a scan rate of 10°C min$^{-1}$. Calorimetric data have been used for the estimation of the Cure Degree (DC) of the samples.

### 2.2.4. Scanning electron microscopy (SEM) Analysis

Micrographs of the carbon nanofibers CNF2500 and T20BD+1.3%CNF2500 nanocomposite were obtained using Scanning Electron Microscope-SEM (mod. LEO 1525, Carl Zeiss SMT AG, Oberkochen, Germany). All samples were placed on a carbon tab previously stuck to an aluminum stub (Agar Scientific, Stansted, UK) and were covered with a 250 Å-thick gold film using a sputter coater (Agar mod. 108 A). Nanofilled sample sections were cut from solid samples by a sledge microtome. These slices were etched before the observation by SEM. The etching reagent was prepared by stirring 1.0 g potassium permanganate in a solution mixture of 95 mL sulfuric acid (95–97%) and 48 mL orthophosphoric acid (85%). The filled resins were immersed into the fresh etching reagent at room temperature and held under agitation for 36 h. Subsequent washings were done using a cold mixture of two parts by volume of concentrated sulfuric acid and seven parts of water. Afterward the samples were washed again with 30% aqueous hydrogen peroxide to remove any manganese dioxide. The samples were finally washed with distilled water and kept under vacuum for 5 days before being subjected to morphological analysis.

### 2.2.5. Tunneling Atomic Force Microscopy (TUNA) Analysis

Atomic force microscope (AFM) images were acquired in an ambient atmosphere (30%-40% humidity) with a Dimension 3100 coupled with a Bruker NanoScope V multimode AFM (Digital Instruments, Santa Barbara, CA) controller operating in tunneling current mode (TUNA-AFM), using microfabricated silicon tips/cantilevers. The sample slices of the T20BD+1.3%CNF2500 were etched before the morphological observation. The TUNA-AFM measurements were performed using platinum-coated probes with nominal spring constants of 35 N m$^{-1}$ and electrically conductive tip of 20 nm. TUNA -AFM operates in contact mode. TUNA works similarly to C-AFM but with higher current sensitivity. The sensor signal is the electric current between the afm tip and the conductive sample for an applied DC bias. In feedback mode, the output signal is the DC bias, adjusted to maintain the electric current setpoint. The following values of the TUNA control parameters are used: DC sample bias ranged from 1 V to 2 V taking into account that bias limit is 12 V, current sensitivity was 1 nA/V, current range was 200 nA, samples/lines: determines the number of data points or pixels in X and Y (256), scan rate: controls the rate at which the cantilever scans across the sample area (0.9-1.5 Hz s-1). In order to obtain repeatable results, different regions of the specimens were scanned. By adopting TUNA-AFM is possible to perform electrical characterization at nanoscale level without grounding the samples. The images were analyzed using the Bruker software Nanoscope Analysis 1.80 (Build R1.126200). Tunneling atomic force microscopy which has been given the official designation 'TUNA' by the equipment manufacturers (Bruker) is a highly sensitive technique by which ultra-low currents (<1pA) ranging from 80 fA to 120 pA can be measured [45]. TUNA allows a tunnelling current to be obtained from a nanosharp tip attached to a cantilever while simultaneously moving the tip across the sample surface to measure topographical data. In contrast to standard STM which requires sample surfaces to be smooth on the nanometer scale, TUNA can investigate surfaces with an r.m.s. roughness of several microns. Moreover, the surface can be studied over scan areas up to hundreds of square microns, allowing a wider picture of the overall morphology to be obtained. Furthermore, unlike the constant-current mode of STM, the physical tracking of TUNA means that the height data collected from the deflection of the cantilever avoids possible artefacts introduced by variations in the conductivity of the sample surface. Another major advantage of TUNA is that it has very high current sensitivity with a noise level of 50 fA [51].

*2.2.5.1. Tunneling Atomic Force Microscopy (TUNA) setup*

The TUNA setup employs a conductive AFM probe, an external voltage source needed to apply a potential difference between the tip and the sample holder, and a current amplifier used to convert the (analogical) current signal into (digital) voltages that can be read by the computer [36,52]. In particular, in our experiments, TUNA operated with a cantilever holder and an epoxy nanofilled sample (containing conductive carbon nanofibers) electrically connected to an external voltage source (see Fig. 2).

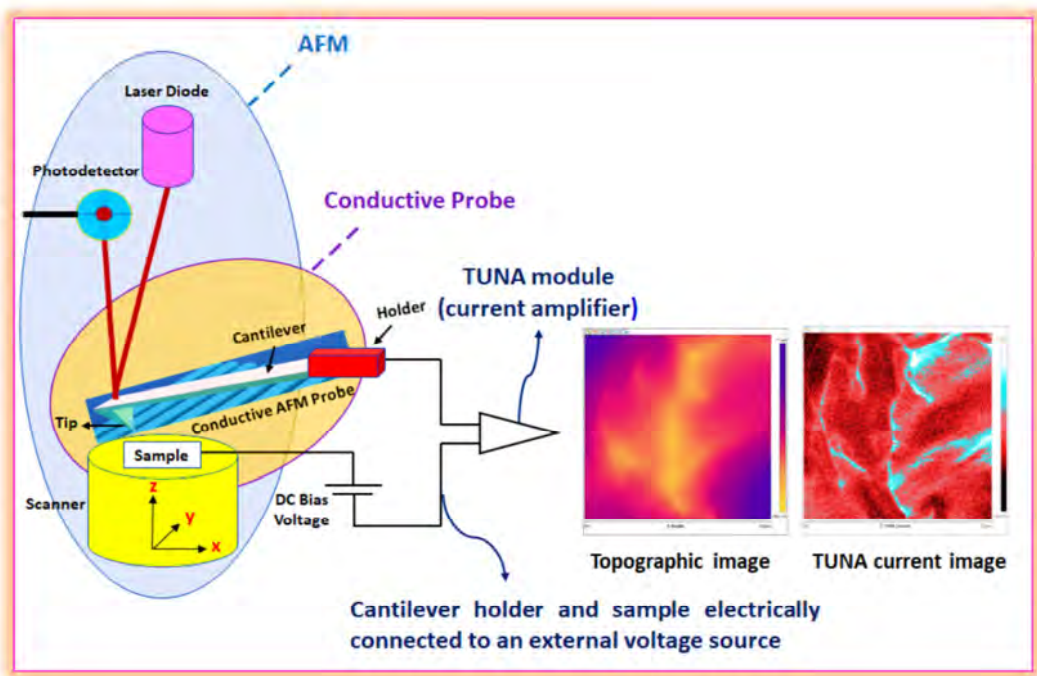

**Fig. 2.** Illustration of Tunneling AFM (TUNA) setup.

In TUNA experiments, the sample is usually fixed on the sample holder using a conductive tape or paste, being silver paints the most widespread. In this regard, it is worth noting that the samples investigated in this work have not undergone any prior treatment with silver paint that is usually used to create electrical contacts to the ground. The principle of the TUNA mode (see Fig. 2), based on an ohmic contact formed between the conductive AFM tip and the sample surface, is the same as C-AFM, by which simultaneous topographic imaging and current imaging can be collected. For the electrical measurements of the samples, just like contact mode AFM, the z-feedback loop uses the dc cantilever deflection as a feedback signal to maintain a constant force between the tip and the sample to generate the topography image [36,53]. In fact, during scanning, a constant DC bias voltage is applied between the tip and the sample and thus the resulting current through the sample is measured with a current amplifier in order to obtain the desired electrical information. TUNA is performed by adding a specialized module to the AFM head and mounting the cantilever in a holder with a current output. The module contains a test connector for calibration and a sensor input connector for connection with the cantilever holder. During operation, the tip-sample force (deflection setpoint) and DC bias voltage are then adjusted to optimize contrast between low and high conductivity regions on the surface [54]. In this work, the TUNA application module was operated in imaging mode, where images of the electrical current are obtained. The amplification is especially important for samples with high resistance or even isolating behavior, where the measured currents can be as low as several femto- to picoamperes. The TUNA technique bears a striking resemblance to Scanning Tunneling Microscopy (STM) using similar technical solutions for the current signal acquisition. However, there is a fundamental difference in the operational principle. In contrast to STM, the TUNA measures the current signal completely independent from the topography which is simultaneously recorded via the cantilever deflection. In order to obtain a potential drop and current across the sample, the external bias is applied to the sample. It is also worth to notice that the contact to the sample should be preferentially ohmic, otherwise the influence of the additional barrier has to be considered carefully. The commonly used range of the applied voltage is $\pm$ 10 V. The application of higher voltages is possible using external voltage sources, though additional circuit protection should be implemented [36].

*2.2.6. Electrical characterization*

The measurements of electrical properties have been carried out both under dc and ac voltage. The DC volume conductivity of the composites have been performed by using disk-shaped specimens of about 2 mm of thickness and 50 mm of diameter. Before performing the electrical measurements, the samples are thermally pre-treated at 90°C for 24 h in order to remove any trace of moisture or solvents. Then, both sides of the samples have been metallized (circular form of about 22 mm of diameter) with silver paint (Alpha Silver Coated Copper Compound Screening, with a resistivity of 0.7 $\Omega$-square) in order to reduce the effects of surface roughness and to ensure Ohmic contacts. The measurement system is composed by a multimeter Keithley 6517A with function of voltage generator (max $\pm$ 1000 V) and voltmeter (max $\pm$ 200 V) and an ammeter HP34401A (having resolution of 10nA). The AC properties have been evaluated in the frequency range 100 Hz–1MHz by using a Quadtec7600 dielectric analyzer and by applying a sinusoidal stimulus of amplitude 0.1V or 1V for specimens above or below the electrical percolation threshold (EPT), respectively, in order to avoid measurements around the current saturation limit (100 mA) of the instrument for samples with low resistivity. Two tests were performed for each composition. All electrical measurements were carried out at room temperature.

## 3. Results and Discussion

*3.1. FTIR analysis of curing reaction*

The epoxide functional group is the characteristic group in the epoxy resin, and epoxide rings will open under the attack of amine molecule during the curing reaction of epoxy resins by amine curing agent, which will decrease the content of epoxide groups drastically. FTIR spectroscopy is an effective method to investigate the curing behavior of epoxy resin by determining the band intensity change of functional groups, before and after curing reactions of epoxy resin and also during the reaction time at a given temperature. We followed the absorbance of the IR peak at 906 cm$^{-1}$ due to the epoxy group responsible of the polymerization reaction. The absorption intensity of this band (906 cm$^{-1}$) is considered maximum for uncured epoxy resins, decreasing with the conversion of the epoxide groups during the curing cycle, when the conversion of the epoxide groups takes place. FTIR spectra allowed us to confirm the complete curing for all the epoxy formulations cured up to 200°C for 3 h, both unfilled resin T20BD and epoxy nanocomposites loaded with different percentages by weight of CNF2500 nanoparticles. FTIR spectra of the unfilled epoxy resin T20BD and CNF epoxy nanocomposites cured at room temperature, up to 125°C for 1h and up to 200°C for 3 h are shown in Figs. 3-4-5 respectively. From the direct comparison between the different FTIR profiles of the epoxy formulations, before and after curing, it is possible to monitor the crosslinking progress allowing to obtain useful information about the chemical reactions that lead to the formation of a highly crosslinked 3D network. In this work, curing mechanism of TGMDA epoxy resin in presence of an aromatic primary diamine DDS curing agent, whose high reactivity is attributed to the high nucleophilicity of the nitrogen atom of the amino group, was evaluated by FTIR. The amino group shows well defined absorptions. Although the N-H stretching is located between 3500 and 3300 cm$^{-1}$, primary amines show a doublet (reflecting the symmetric and antisymmetric stretching modes), while the secondary amines show one single band. It can be seen from FTIR profiles that at room temperature (see Fig. 3) the relative absorption intensity of epoxide group still remains obviously, when the curing temperature was up to 125°C for 1h (see Fig. 4), the absorption intensity at 906 cm$^{-1}$ weakened compared to that cured at room temperature. It is worth pointing out that the curing time had a little influence on the opening reaction of epoxide ring compared with 1 h cured. FTIR spectra of the unfilled epoxy resin T20BD and CNF epoxy nanocomposites cured up to 200°C for 3 h (see Fig. 5) show that both the asymmetric ring stretching band of the epoxy ring (906 cm$^{-1}$) and the N-H stretching vibration bands of DDS at 3060-3500 cm$^{-1}$ disappeared as the reaction proceeded while the absorbing bands of the O-H stretch (produced by the attack of NH$_2$ to the epoxide ring) at 3200-3650 cm$^{-1}$ increased. At the end of the curing, the C-O-C band of the epoxy ring and N-H stretch bands have almost completely disappeared. The absence of the absorption of epoxy ring and presence of OH group confirm the conversion of epoxy group into the corresponding polymer and thus that all the amine and epoxy groups reacted during the curing process proving undoubtedly the proper formation of the crosslinked network. It is worth noting that the changes in the intensity of the peaks mentioned above are clearly visible compared to the peaks at 1143 cm$^{-1}$ and 1105 cm$^{-1}$ corresponding to the strong asymmetric and symmetric SO$_2$ stretching of the DDS hardener [55] and the peaks at 1595 cm$^{-1}$ and 1515 cm$^{-1}$ attributable to the phenyl groups [56], which remain unchanged during the curing process.

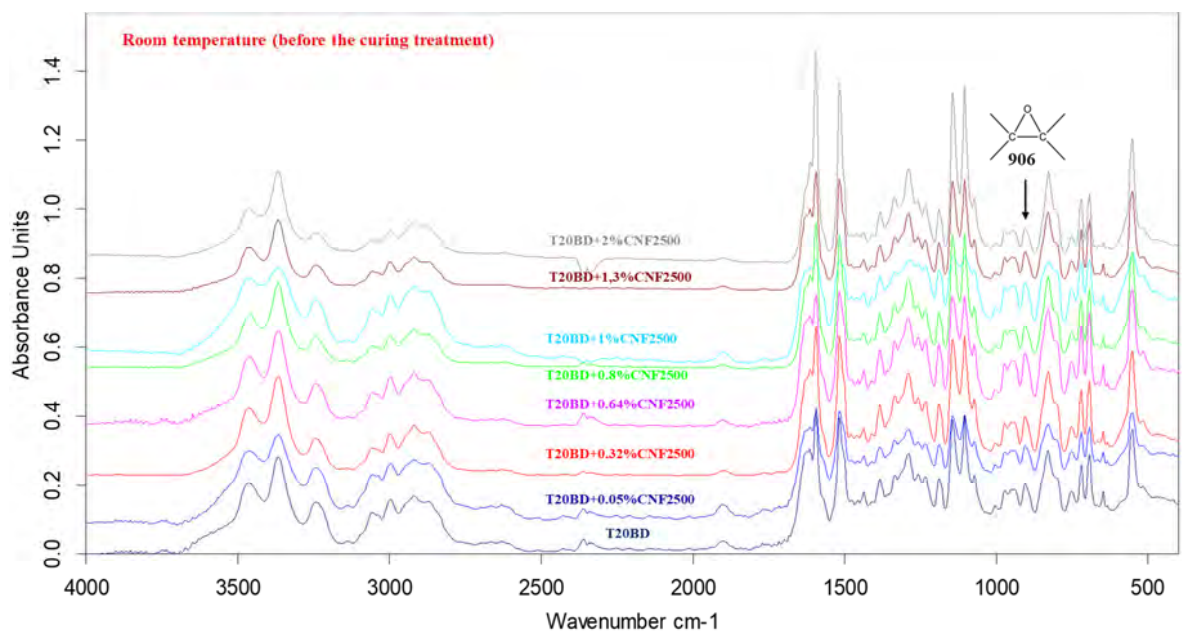

**Fig. 3.** FTIR spectra of the unfilled epoxy resin T20BD and CNF epoxy nanocomposites at room temperature before the curing treatment.

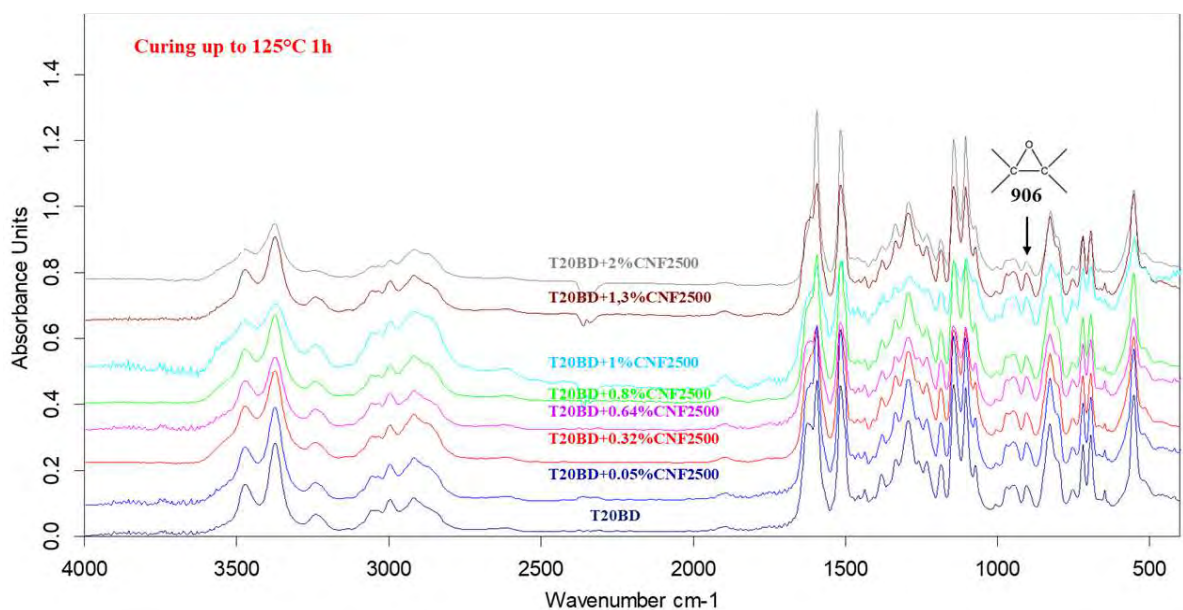

**Fig. 4.** FTIR spectra of the unfilled epoxy resin T20BD and CNF epoxy nanocomposites cured up to 125°C for 1 h.

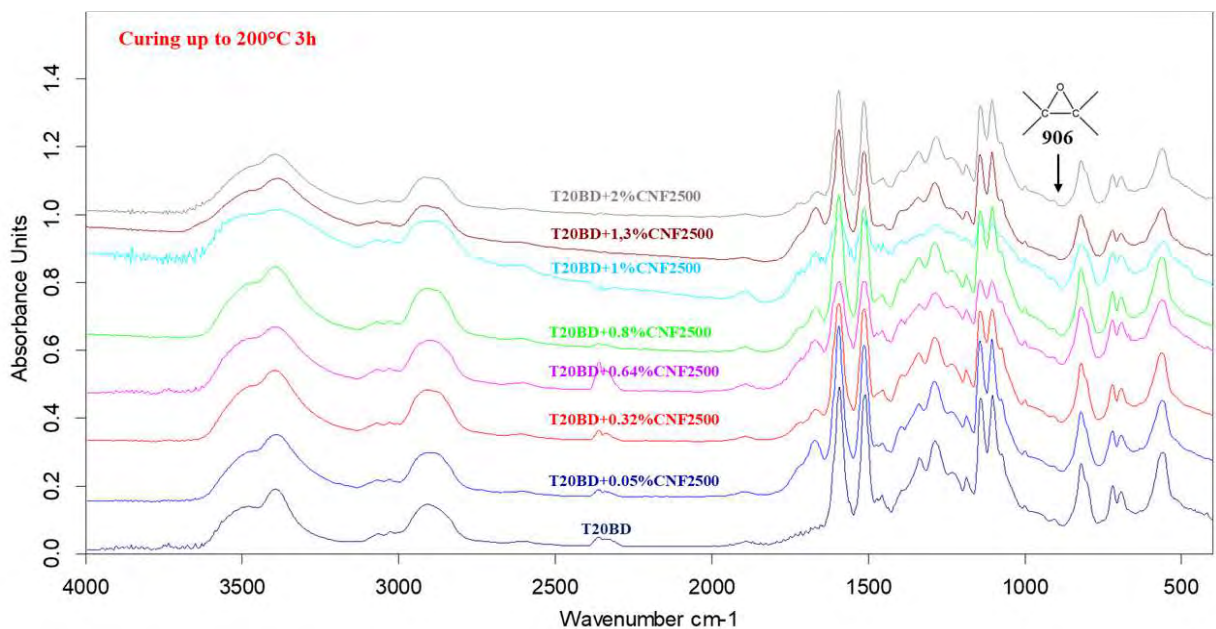

**Fig. 5.** FTIR spectra of the unfilled epoxy resin T20BD and CNF epoxy nanocomposites cured up to 200°C for 3 h.

### 3.2. Differential Scanning Calorimetry (DSC) Analysis

In this work, DSC has been used for the estimation of the curing degree (DC) of the samples under the assumption that the exothermic heat evolved during cure is proportional to the extent of reaction. The DC can be determined from the total heat of reaction ($\Delta H_T$) of the curing reaction and the residual heat of reaction ($\Delta H_{resid}$) of the partially cured epoxy resin [3] as follows:

$$DC = \frac{\Delta H_T - \Delta H_{resid}}{\Delta H_T} \times 100 \qquad (1)$$

To obtain fraction reacted at various temperatures, we have performed a series of isothermal experiments. To secure accurate total $\Delta H_T$ values from isothermal studies, dynamic runs were made after the isothermal curing cycle to obtain the residual heat of reaction. The total heat of reaction was considered as follow:

$$\Delta H_T = \Delta H_{iso} + \Delta H_{resid} \qquad (2)$$

where $\Delta H_{iso}$ and $\Delta H_{resid}$ are the areas under the isothermal and dynamic thermograms, respectively.

In this paper, the curing degree (DC) has been obtained in dynamic and isothermal regime. This last experimental procedure has also been considered because many processes in the industry are actually done isothermally, the curing degree obtained in dynamic regime has been compared with the curing degree of the samples after a curing cycle composed of two steps: a first step of 125°C for 1 hour followed by a second step at the higher temperature of 200°C for 3 hours, where both the steps have been carried out in isothermal condition in oven. In particular, in the dynamic regime, the uncured nanofilled samples were analyzed by a three step dynamic heating program in the temperature range between -50°C and 300°C considering a first run from -50°C up to 300°C with a scan rate of 10°C min$^{-1}$, a second run from 300°C to -50°C with a scan rate of 50°C/min and a third run from -50°C up to 300°C with a scan rate of 10°C/min (see Figs. 6a-b).

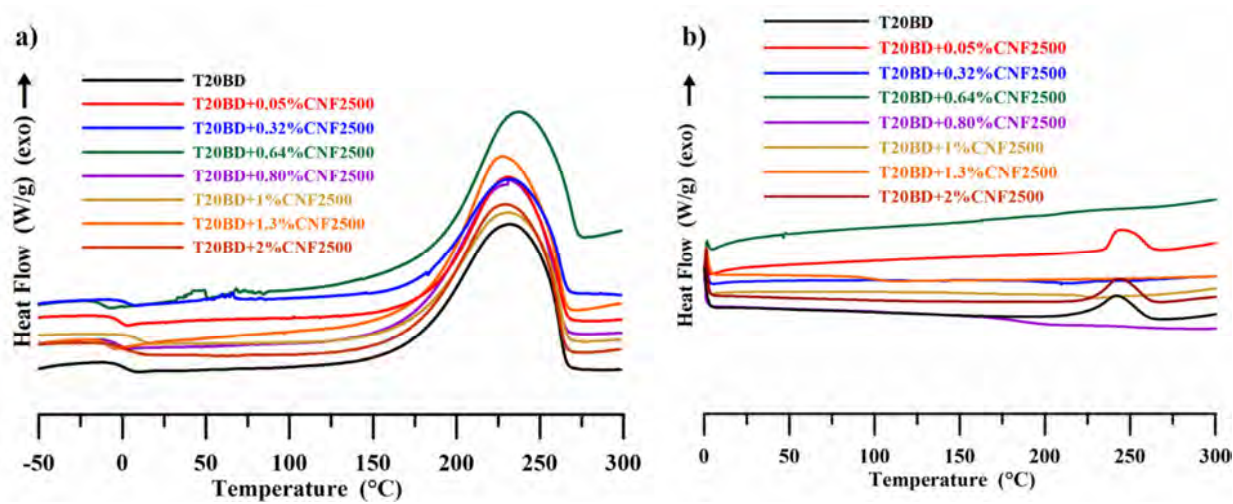

**Fig. 6.** DSC curves of the nanofilled samples: a) in the first run, b) in the second run.

The samples oven cured at 200°C were analyzed by a single heating run from 0°C up to 300°C with a scan rate of 10°C/min. The DC was determined from the total heat of reaction ($\Delta H_T$) of the first run curing reaction and the residual heat of reaction ($\Delta H_{resid}$) of the partially cured epoxy resin as above decribed. The results of calorimetric analysis performed using the two different experimental procedures are shown in Fig. 7.

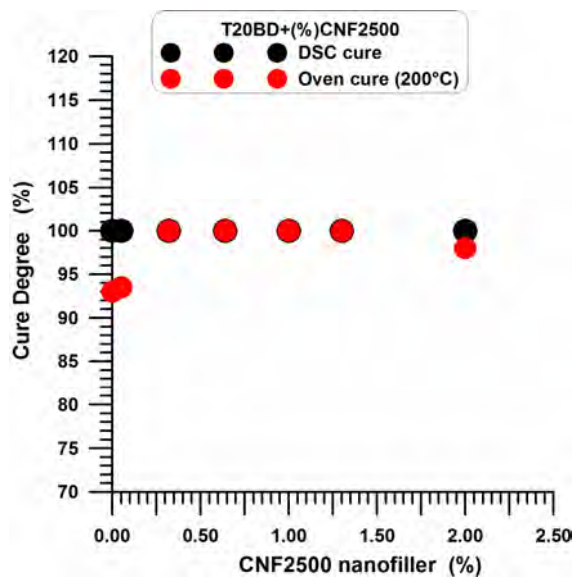

**Fig. 7.** DC values of the CNF epoxy nanocomposites cured under dynamic and isothermal heating conditions.

All the analyzed formulations show DC values higher than 92% also in isothermal regime. The nanofiller causes an increase in the efficiency of curing process in isothermal regime, which is usually the real condition of curing in industrial processes. The DC of all samples containing embedded CNFs is very high compared to unfilled epoxy resin and all CNFs filled epoxy samples reach a value of almost 100%. The results on the resin filled with CNF nanoparticles highlight an intersting behaviour with respect to the curing degree, in fact it meets the industrial requirements in many fields related to the application of structual functional resins (aeronautics, wind energy etc..). Furthermore, it has been found that all CNF epoxy nanocomposites are more thermally stable compared to unfilled epoxy matrix T20BD. In fact, thermogravimetric analysis highlighted for the carbon nanofibers-reinforced resins a stabilizing effect of the carbon nanofibers CNF2500 in the first stage of the thermal degradation process [4].

### 3.3. Dynamic Mechanical Analysis (DMA)

Dynamical mechanical data provide useful information on the relaxation processes that become operative in the polymer in a temperature range depending on the examined system. Figs. 8 and 9 show the storage modulus and loss factor (tan$\delta$) of the T20BD unfilled epoxy resin and epoxy nanocomposites loaded with carbon nanofibers, respectively. A decrease in the storage modulus with respect to the unfilled resin T20BD in the temperature range -50°C to about 250°C is observed for T20BD+0.05%CNF2500 sample; although the detected values are suitable in the usual operational temperature range of structural materials. The lower value in the storage modulus of the epoxy sample loaded with 0.05% by wt of CNF2500 is most likely due to a lower crosslinking density. The loss factor (tan $\delta$) is a measurement of damping property, which is the relation between the elastic energy stored and the energy dissipated per cycle of vibration. The glass transition temperature (Tg) for the T20BD+0.05%CNF2500 sample decreases with respect to the epoxy formulation T20BD. The decrease observed in the temperature of the main peak of tan $\delta$ curve is most likely due to a lower density of the resin network that leads to easiest chain motion. We can observe for the neat resin

T20BD and T20BD+0.8%CNF2500 nanocomposite almost the same trend in the storage modulus in the entire analyzed temperature range -90°C to about 260°C. As expected, no significant changes are observed in the glass transition temperature (Tg) for the formulations T20BD and T20BD+0.8%CNF2500, suggesting that the nanofiller content of 0.8% by wt does not hinder the motion of polymer segments. For the epoxy nanocomposites loaded with percentages of 1% and 1.3% of CNF2500, we can observe a general increase in the storage modulus with respect to the unfilled resin T20BD, which however becomes much more marked especially in the range between -90°C to about 60°C. In the epoxy nanocomposites T20BD+1%CNF2500 and T20BD+1.3%CNF2500, the carbon nanofibers CNF2500 show a reinforcing effect up to 60°C. In any case, the fact that the Tg values are in the range between 260°C and 270°C and the value in the storage modulus is higher than 2000 MPa up to 110°C for all the analyzed samples, lead to the conclusion that the formulated carbon nanofibers reinforced resins are suitable for a very wide range of structural applications.

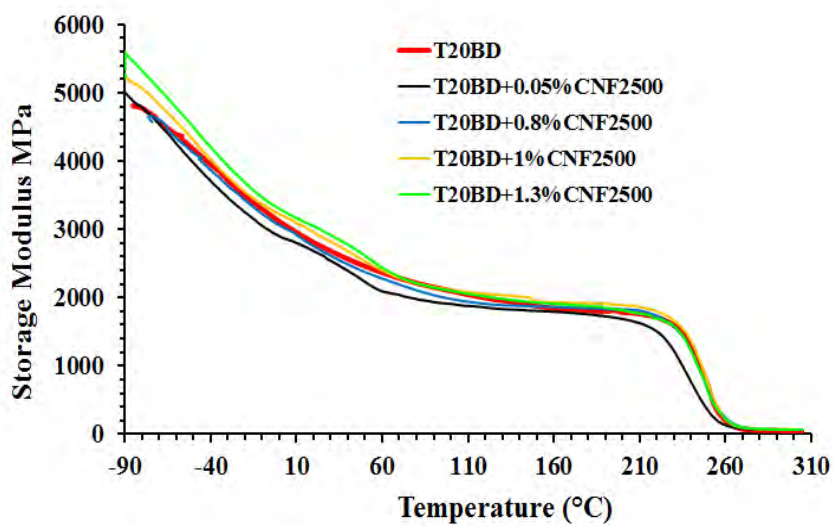

**Fig. 8.** Storage modulus of the T20BD unfilled epoxy resin and epoxy nanocomposites loaded with carbon nanofibers.

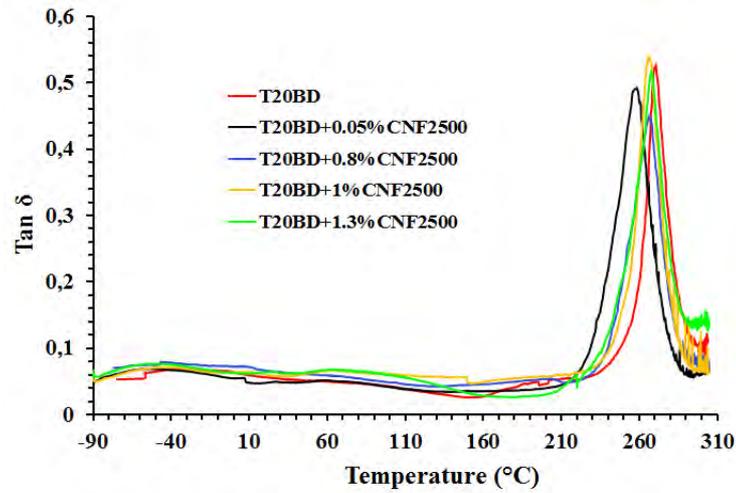

**Fig. 9.** Loss factor (tanδ) of the T20BD unfilled epoxy resin and epoxy nanocomposites loaded with carbon nanofibers.

### 3.4. DC Electrical Properties

A detailed study of the DC electrical properties of the composites has been reported in a previous paper [3]. For sake of clarity and completeness of the present work, the main observed features are here briefly summarized. Fig. 10 shows the DC volume conductivity of the composite as a function of filler loading (wt%).

The electrical percolation threshold (i.e. EPT) falls in the narrow range [0.05÷0.32] wt%. The composite below the EPT (i.e. 0.05 wt%) presents an electrical conductivity of the order of pS/m comparable with that of neat resin. Above the electrical percolation threshold, due to the formation of an electrical network of neighboring conductive particles and the possibility for electrons to flow through tunneling effect, the composites present a transition of the electrical behaviour from an insulating to a conductive one [57]. Therefore, the electrical conductivity of the composites increases significantly above the EPT following a power-law of the form:

$$\sigma = \sigma_0 (\phi - \phi_c)^t \qquad (3)$$

where $\sigma_0$ is the intrinsic conductivity of the filler, $\phi_c$ is the *EPT* and $t$ is a critical exponent depending on the dimensionality of the percolating structure [58]. At the highest filler concentration (i.e. 2 wt%) the conductivity achieves the value of 2 S/m. The characteristic parameters of the eq. 3 can be estimated by reporting the log–log plot of the electrical conductivity as function of the filler amount, as shown in the inset a) of the Fig. 10. More in details, the value for the critical exponent $t$ can be obtained as the slope of the fitting curve. Instead, the value of the estimated electrical percolation threshold (i.e. $\phi_{est.}$) is used as a fit parameter until the maximum of the regression coefficient of the interpolating line is achieved (i.e. $R^2$ close to 1). The value of 2.2 for the exponent $t$ and 0.32 wt% for $\phi_{est}$, are obtained by means of this procedure. These figures are in agreement, respectively, with universal values observed for composites filled with conductive nanoparticles forming a continuous 3-dimensional percolating paths throughout the matrix and with the percolation threshold detected experimentally.

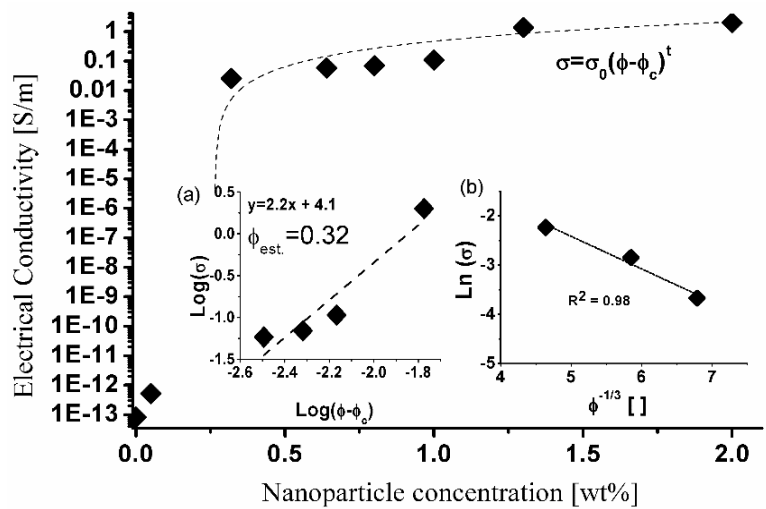

**Fig. 10.** Conductivity of nanocomposite systems as a function of the fillers concentrations (wt%).

*3.5. AC Electrical Properties*

Electrical properties of insulating polymers can be improved by adding high-permittivity nanofillers such as carbon-based particles [59]. Such materials are promising candidates for novel applications in the electromagnetic (EM) field for use as EM coatings, shields, filters and radar absorber materials (RAMs) [60]. Since the influence of filler parameters on the electrical properties are still unclear, the experimental investigation in the frequency domain may be useful in order to evaluate their effectiveness for such purposes in an attempt to design tailored nanocomposites with specific multifunctionality. A classical technique adopted to this aim is the impedance spectroscopy (IS) based on measurement and subsequent evaluation of the electrical impedance or other related parameters, such as permittivity and electrical conductivity, as a function of the frequency (*f*) of a sinusoidal stimulus, as shown in Figs. 11 and 12.

In particular, Fig. 11a and Fig. 11b report the electrical conductivity as function of frequency in the range [100 Hz ÷ 1 MHz] and versus the nanoparticle concentration [wt%] evaluated at the two frequency extremes (i.e. 100Hz and 1MHz), respectively. It is possible to note that for the composites below the EPT (i.e. T20BD+0.05% CNF2500), the electrical conductivity shows an evident frequency dependent behaviour similar to that observed for insulating matrix (i.e. TGMDA – Pure Resin) with a progressive increase for higher frequency. On the contrary, for all other formulations that are above the percolation threshold the conductivity is frequency-independent in almost the analyzed frequency range [100Hz-1MHz]. It is well known in fact that, starting from a certain critical frequency $f_c$, the AC conductivity increases according to the so-called Almond-West-type power law [61,62]:

$$\sigma(\omega) = \sigma_{DC} + A \cdot \omega^s \qquad (4)$$

where $\sigma_{DC}$ is the DC conductivity, $A \cdot \omega^s$ is the AC conductivity. More in details, $A$ denotes a temperature-dependent constant, $\omega$ is the angular frequency (i.e. $\omega = 2\pi f$) and $s$ is a characteristic exponent whose value is strictly close to 1. Moreover, with reference to Figure 11b, it is possible to observe how the percolation curve evaluated at low frequency (i.e. 100Hz) is very similar to that one identified with DC measurements. In addition, the percolation curve evaluated at f=1MHz overlaps well with that measured at 100 Hz except for the insulating composites due to frequency dependent behavior of the electrical conductivity observed in Fig. 11a.

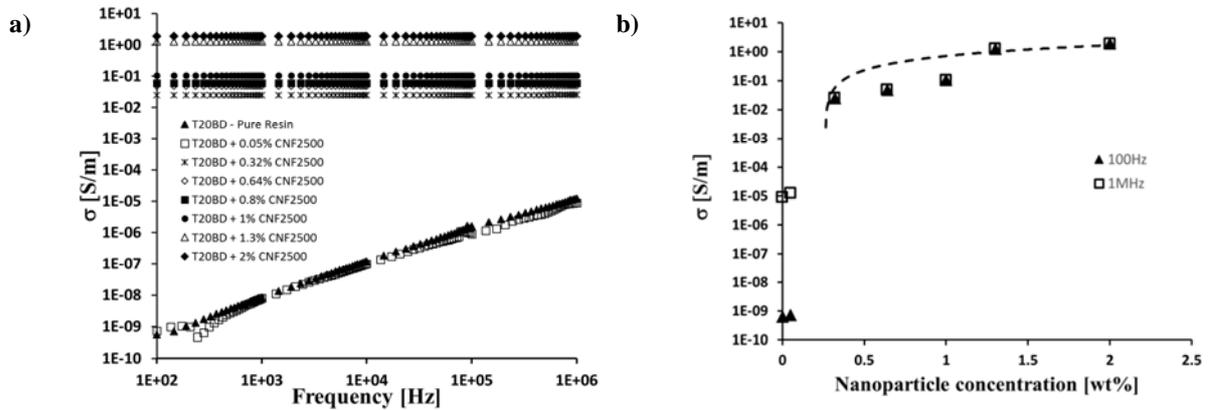

**Fig.11.** Conductivity of nanocomposite systems as a function of the frequency (a) and as function of the filler loading evaluated at the frequency of 100Hz and 1MHz (b).

An interesting property that plays a key role in frequency-dependent applications of carbon-based nanocomposites, is the complex effective (relative) permittivity. It describes the polarization of a material under an applied variable electric field:

$$\varepsilon^*(\omega) = \varepsilon' - j \, \varepsilon'' \tag{5}$$

where $j^2 = -1$, $\varepsilon'$ is the relative permittivity or dielectric constant which takes into account the stored energy in a material subject to an external variable field, whereas $\varepsilon''$ is a term which, for a dielectric material, provides an information on the efficiency to dissipate the power of the applied electromagnetic field. The larger its value, the greater the capacity of absorbing the incident electromagnetic field that is dissipated as heat.

The variations of the real part of the dielectric permittivity against the frequency is presented in Fig. 12a. The effect of temperature rise could be significant due to the greater freedom of the dipole molecular chains within the epoxy. However, since such electrical characterization is carried out at constant temperature, it can be neglected. The behaviour of the dielectric constant can be associated to the presence of free dipolar functional groups (mainly of type C-OH or sometimes N-H as reaction product) and/or an interfacial polarization attributable to the presence of conducting impurities. Results show that, regardless of the formulation of the samples, the value of $\varepsilon'$ slightly decreases with increasing frequency, which suggests that the main contribution to the polarization comes from orientation mechanisms. Moreover, the dielectric permittivity depends exponentially on the carbon nanofiller content, as reported in Fig. 12b and already observed in other literature studies [63-64]. This increase of ε' in the polymer nanocomposites caused by the growing percentage of conductive particles can be attributed to the enhancement of interfacial polarization [65]. Dielectric materials with high dielectric permittivity are particularly required in several industrial fields such as modern electronics, electrical power systems, hybrid electric vehicles, and so on. Neat polymers are the best choice of dielectric materials for charge storage applications, thanks to their high breakdown strength. However, their potential applications are restricted by the typical low value of dielectric permittivity. Therefore, polymer nanocomposites have shown great potential in overcoming such limitation since the introduction of small amount of nanofillers can increase ε of resulting materials without sacrificing the dielectric strength and other interesting mechanical, thermal and chemical properties [66]. Finally for completeness of analysis, Figs. 12c 12d show the evolution of imaginary part (i.e. ε'') of the complex permittivity as a function of the frequency and as function of the filler loading evaluated at 100Hz, respectively. For clarity of graph, Fig.12c illustrates only the behavior for the unfilled resin (i.e. T20BD – Pure Resin) and for the composition at the highest filler concentration (i.e. T20BD+2%CNF2500). The parameter ε'' is linked to the electrical conductivity from the following analytic relationship:

$$\varepsilon''(\omega) = \frac{\sigma}{\omega \cdot \varepsilon_0} \qquad (6)$$

where $\varepsilon_0$ is the vacuum permittivity (8.856 pF/m). I may be noted that, for composites above the EPT having an almost constant conductivity in the frequency range under examination, the value of ε'' decreases linearly with the angular frequency ω. Vice versa for the pure resin (but it is also valid for the formulation T20BD+0.05%CNF2500) the imaginary part of the complex permittivity shows a constant trend.

Furthermore, as evident from Fig. 12d and due to (6) the evolution of the ε'' against the nanoparticle concentration [wt%] follows a trend similar to that of electrical conductivity. This confirms the validity of the following equation, already adopted for the electrical conductivity (eq. 3), useful to describe the changes of the electrical properties *p* (i.e. ε, σ, etc.) of composites above the percolation threshold $\phi_c$:

$$p \propto |(\phi - \phi c)|^{\pm b} \qquad (7)$$

where *b* is a critical exponent that is different for various properties [58]. The strong dependence of the properties on the frequency around the percolation threshold allows to suitably use the nanocomposites as a highly sensitive device for monitoring strain in aeronautic structural parts or even as part of wing de-icing system [67-68].

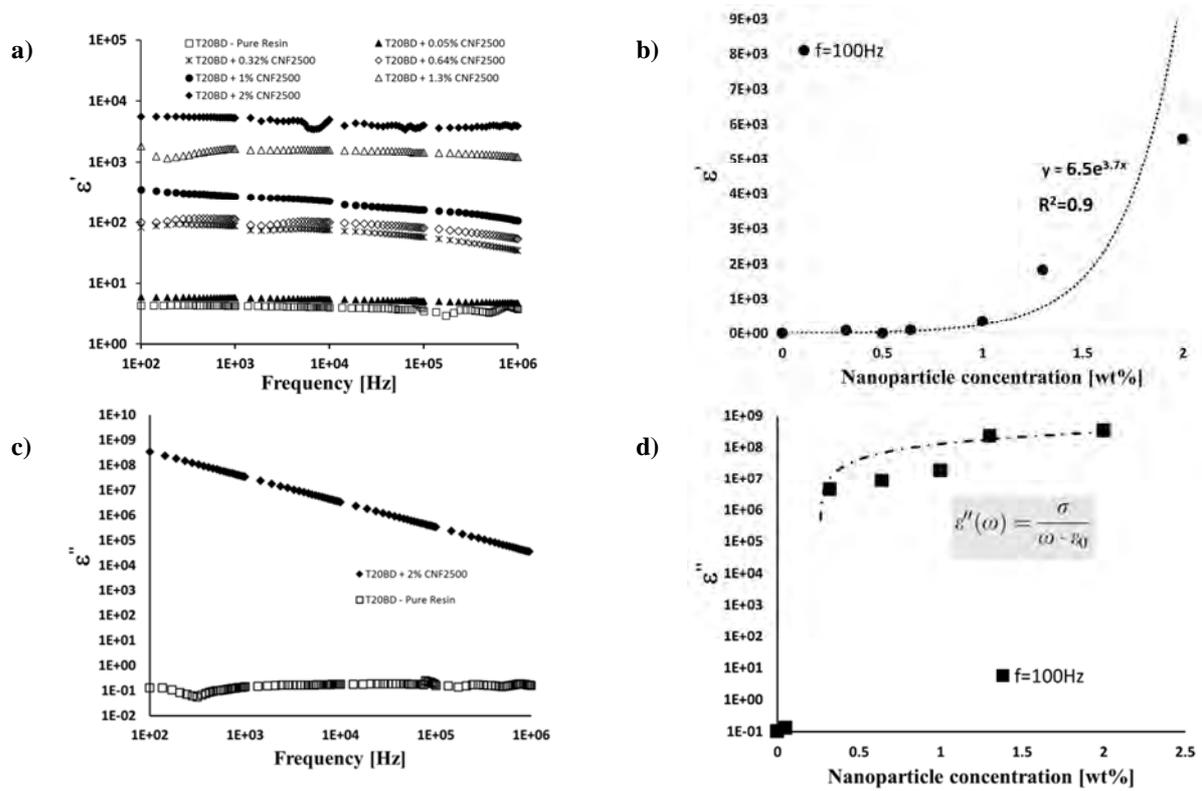

**Fig. 12.** Real part (i.e. ε') and imaginary part (i.e. ε'') of the complex permittivity as a function of the frequency and as function of the filler loading evaluated at the frequency of 100Hz, in *a*, *b*, *c* and *d*, respectively.

### 3.6. Morphological analysis

In a previous paper, TEM investigation on the nanofibers was performed to analyze the morphology of the nanofibers before their incorporation into the epoxy mixture. It was found that heat-treated carbon nanofibers CNF2500 are characterized by straighter walls where the nested configuration characteristic of the as made CNF is not clearly visible [4,35]. In this paper SEM and TUNA images have been recorded for all investigated concentrations of the developed nanocomposites. Fig. 13 shows the SEM image of carbon nanofibers CNF2500.

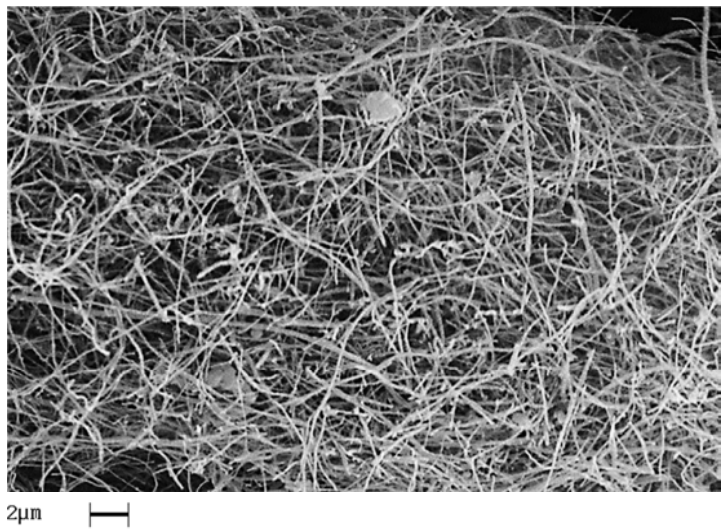

2μm

**Fig. 13.** SEM image of carbon nanofibers CNF2500.

In order to analyze the homogeneity of the nanofiller dispersion in the polymeric matrix, the epoxy samples loaded with CNF2500 were investigated by means of SEM. The analysis was carried out on etched samples to remove the resin surrounding the nanofibers, leaving them bare as described in the experimental section. Fig. 14 shows SEM image of the fracture surface of nanofilled resin at a loading rate of 1.3% by weight. Careful observation allows to detect a homogeneous structure for the T20BD+1.3%CNF2500 sample, in which the CNF2500 are uniformly distributed in the epoxy matrix also highlighting the conductive network of carbon nanofibers strongly anchored to the epoxy matrix. This explains well the good dynamic mechanical properties and the electrical conductivity value of 1.37 S/m observed for this sample. It is worth noting that, in the dark zone of the SEM image, the etching procedure has enabled to better highlight the presence of carbon nanofibers that form a continuous network in the matrix.

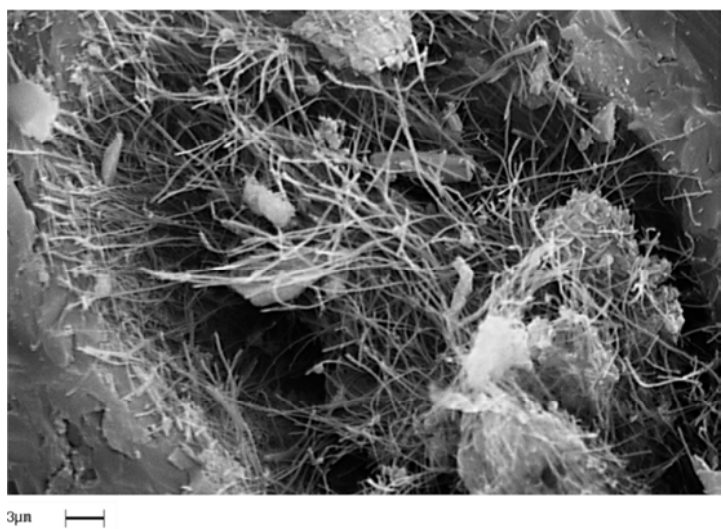

3μm

**Fig. 14.** SEM image of the fracture surface of T20BD+1.3%CNF2500 sample.

Fig. 15 show the TUNA-AFM micrographs and the corresponding 3D profiles of the T20BD+1.3%CNF2500 epoxy formulation. The TUNA-AFM images were collected on etched samples to partially remove the resin surrounding the carbon nanofibers and to better observe the distribution of the nanofiller inside the epoxy matrix. For each sample, 3 TUNA-AFM image types, which are the most common in contact mode, namely height (or topography) image, deflection error image, and tuna current image, are displayed below. The height image is the type of image most commonly reported and published. Usually, the height image is a map of differently coloured pixels, with a colour bar relating the colour to the height. This type of acquisition is really very useful as it allows to estimate both lateral (xy) and height (z) measurements. However, one reason other types of image are here shown is that such "height maps" do not always really "look like" the object in question, in other words, the appearance of a certain shape can be very different to that it would have in optical (or electron microscopy). What this means, is that, to the casual observer, such images do not display easily the shape of the features. Ways around this include shading the image, and more commonly, creating a pseudo-3D image from the height data. Because the deflection error image is equivalent to a map of the slope of the sample, it often displays the shape of the sample more easily. But it is worth noting that the z-scale in deflection is completely meaningless in terms of the sample structure. All it shows is how the tip deflected as it encountered sample topography. It is important to remember too that the best images are obtained when the deflection signals are minimised. In general, TUNA-AFM by simultaneously mapping the topography and current distribution at nanoscale level allows to obtain the direct correlation of a sample location with its electrical properties. In fact, the 3 TUNA-AFM images reported below clearly show the morphological characteristics of the nanofilled sample and provide complementary information that helps the reader to a more complete understanding of the observed electrical properties. In the TUNA-AFM images of the nanofilled sample, the conductive network of carbon nanofibers strongly anchored to the epoxy matrix can be clearly observed, especially in 2D and 3D tuna current images. In fact, the carbon nanofibers appear homogeneously distributed across the sample surface and they are clearly observable in almost any image, but they are most resolved in the tuna current images that have been recorded at the same time as the height and deflection error images. The tuna current images show an increased contrast in the morphological features of carbon nanofibers. A clear correlation between the topography and the regions of high current was found whenever a measurable current can be recorded. In the case of the TUNA current image obtained at a bias voltage within a range of 1 to 2 V, conducting CNF2500 nanoparticles appear very bright in the light blue colour, thus demonstrating their high conductivity and domains with lower conductivity values appear darker. As can be seen in the current profile, domains with different brightnesses present differences in the current value. A careful observation of the current profile of the T20BD+1.3%CNF2500 sample (see TUNA current image in Fig. 15) allows to confirm that the carbon nanofiber reinforced resins are intrinsically conductive. In fact, for the sample T20BD+1.3%CNF2500, currents ranging from 122.8 fA to 1.1 pA were detected. The possibility to detect such low currents (ranging from fA to pA) in the nanocomposite confirms the relative high value of the electrical conductivity (1.37 S/m) and the effective conductive paths due to an optimal conductive nanofiller dispersion as it is highlighted by the strong contrast of the colors in the tuna current micrographs. Current flow through tunneling effect along the carbon nanofibers ensures a good transfer of electrical properties to the polymeric surface through a conductive network at nanoscale level. The presence of conductive network inside the epoxy resin is clearly visible in Fig. 16. In particular, the distribution and characteristics of the network are directly observed by TUNA current images (bidimensional and 3D profiles) where very conductive regions due to the presence of brighter CNF2500 filaments are detected on the surface of the nanocomposite.

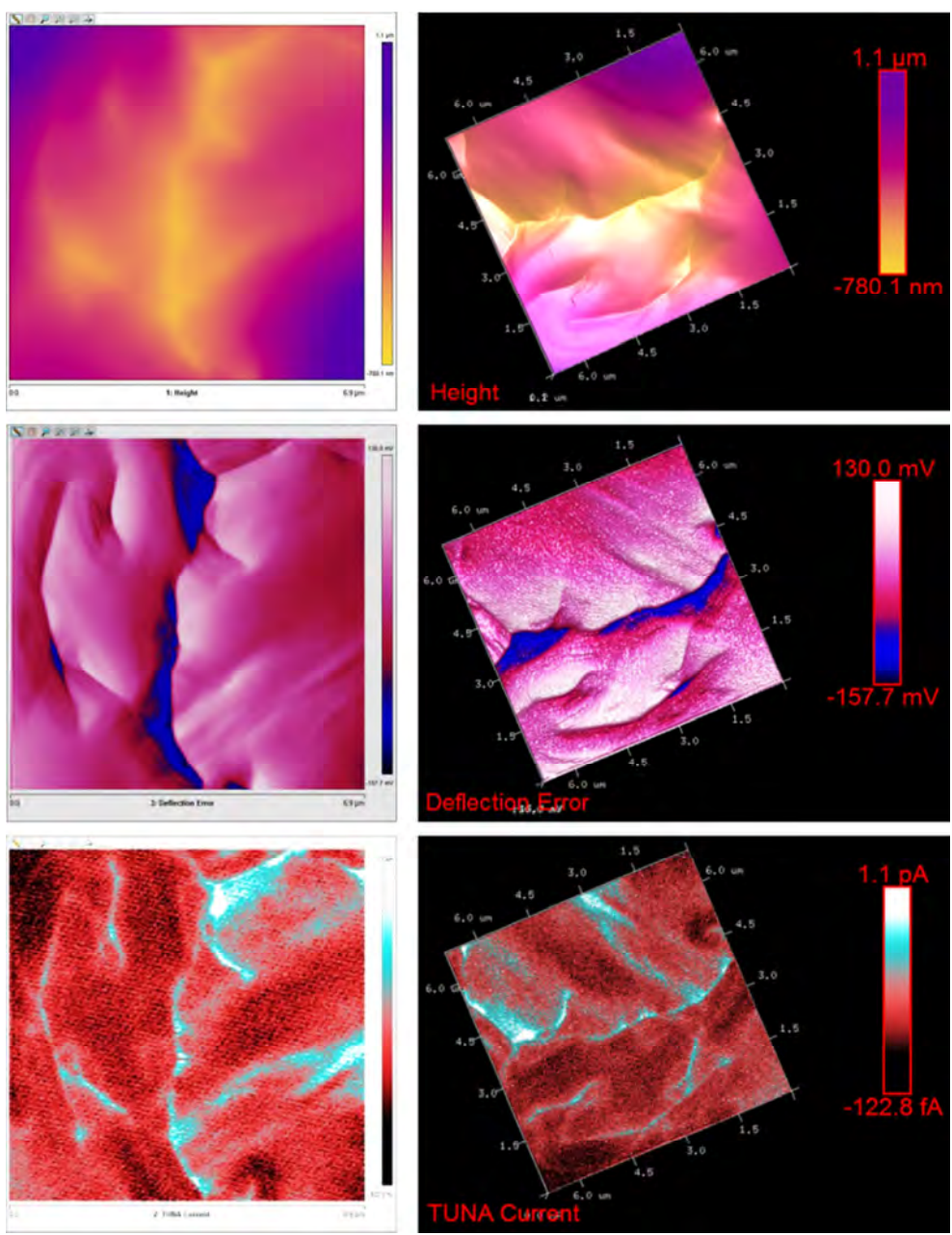

**Fig. 15.** TUNA-AFM micrographs (6.9 μm × 6.9 μm) of the fracture surface of the T20BD+1.3%CNF2500 sample (see the 3 images from top to bottom on the left side: height, deflection error, and tuna current images) and the corresponding 3D profile (see the 3 images from top to bottom on the right side: height, deflection error, and tuna current images).

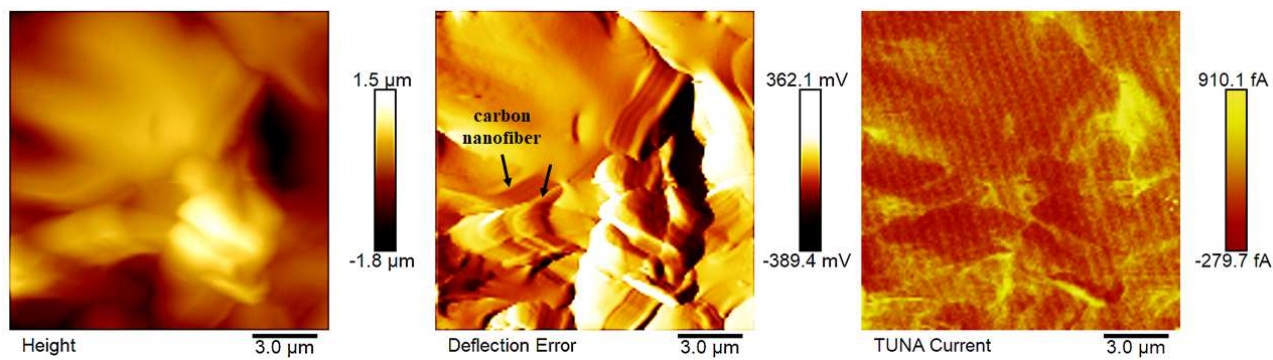

**Fig. 16.** TUNA-AFM height, deflection error, and tuna current images of the fracture surface of the T20BD+1.3%CNF2500 sample.

## 4. Conclusion

The TUNA-AFM technique has proven to be really a powerful tool for measuring electric transport through the carbon nanofiber based epoxy network, thus allowing to provide a map of nanofiller distribution at atomic scale level inside the polymeric matrix. In the analyzed nanocomposites, the carbon nanofibers are integrated into the matrix and become part of the cross-linked structure. The lower percolation threshold and higher conductivity exhibited by the carbon nanofiber reinforced nanocomposites can be justified on the basis of their stiffness and smoothness of surface graphitized CNF2500 which determines a lower thickness of the insulating epoxy layer around the fibers. When the percolation threshold is reached, the formation of a percolated filler network structure takes place, where a critical minimum distance between carbon nanofibers has been reached and electron conduction is facilitated through a 'tunnelling' mechanism. The TUNA technology is an innovative tool able to correlate the electrical properties of the nanofilled structural material with its topographic features, giving new insights for a better understanding of current conduction mechanism. The use of this technique allowed to identify the nanoscopic distribution of the conductive phase and characteristics of the charge conduction throught the current pathways in carbon nanofiber nanocomposites, without undergoing any prior treatment with silver paint that is usually used to create electrical contacts to the ground, proving that the investigated samples are intrinsically conductive. The CNF/epoxy nanocomposites are characterized by good electrical performance and they are potentially suitable for aircraft lightning strike protection. In particular, for concentration of CNFs higher than 0.25 wt.%, the nanofiller determines an increase in the efficiency of curing process performed in isothermal regime, which is usually the real condition of curing in industrial processes; and the nanocomposites are characterized by an electrical conductivity beyond the EPT and can be used for all applications where the employment of an insulating material is not suitable. In addition, the strong dependence of the properties on the frequency around the percolation threshold allows to suitably use the CNF based nanocomposites as a highly sensitive device for monitoring strain in aeronautic structural parts or even as part of wing de-icing system. The curing behavior, investigated by FTIR and DSC, proving the proper formation of the crosslinked network together with the high mechanical properties demonstrate the real potential of the formulated conductive nanocomposites for advanced structural applications.

### Acknowledgements

The research leading to these results has received funding from the European Union Horizon 2020 Programme under Grant Agreement EU Project 760940-MASTRO.